\documentclass[sigconf]{acmart}
\usepackage{amsmath}
\usepackage{graphicx}
\usepackage{float}
\usepackage{hyperref}
\usepackage{balance}

\AtBeginDocument{%
  }
    
\copyrightyear{2026}
\acmYear{2026}
\setcopyright{cc}
\setcctype{by-nc-nd}
\acmConference[WWW '26]{Proceedings of the ACM Web Conference 2026}{April 13--17, 2026}{Dubai, United Arab Emirates}
\acmBooktitle{Proceedings of the ACM Web Conference 2026 (WWW '26), April 13--17, 2026, Dubai, United Arab Emirates}
\acmPrice{}
\acmDOI{10.1145/3774904.3792950}
\acmISBN{979-8-4007-2307-0/2026/04}

\begin{document}

\title{Real-Time Trend Prediction via Continually-Aligned LLM Query Generation}

\author{Zijing Hui}
\authornote{Corresponding authors}
\affiliation{%
  \institution{Meta Platforms}
  \city{Menlo Park}
  \state{CA}
  \country{USA}
}
\email{zijihui@meta.com}

\author{Wenhan Lyu}
\authornotemark[1] 
\affiliation{%
  \institution{Meta Platforms}
  \city{Menlo Park}
  \state{CA}
  \country{USA}
}
\email{wenhanl@meta.com}

\author{Shusen Wang}
\affiliation{%
  \institution{Meta Platforms}
  \city{Menlo Park}
  \state{CA}
  \country{USA}
}
\email{wangshusen@meta.com}

\author{Li Chen}
\affiliation{%
  \institution{Meta Platforms}
  \city{Menlo Park}
  \state{CA}
  \country{USA}
}
\email{lichen2@meta.com}

\author{Chu Wang}
\affiliation{%
  \institution{Meta Platforms}
  \city{Menlo Park}
  \state{CA}
  \country{USA}
}
\email{wangchu@meta.com}

\begin{abstract}
Trending news detection in low-traffic search environments faces a fundamental cold-start problem, where a lack of query volume prevents systems from identifying emerging or long-tail trends. Existing methods relying on keyword frequency or query spikes are inherently slow and ineffective in these sparse settings, lagging behind real-world shifts in attention. 

We introduce \textbf{RTTP}, a novel \textbf{R}eal-\textbf{T}ime \textbf{T}rending \textbf{P}rediction framework that generates search queries directly from news content instead of waiting for users to issue them. RTTP leverages a \textbf{continual learning LLM (CL-LLM)} that converts posts into search-style queries and scores them using \textbf{engagement strength + creator authority}, enabling early trend surfacing before search volume forms. To ensure adaptation without degrading reasoning, we propose \textbf{Mix-Policy DPO}, a new preference-based continual learning approach that combines on-policy stability with off-policy novelty to mitigate catastrophic forgetting during model upgrades.

Deployed at production scale on Facebook and Meta AI products, RTTP delivers \textbf{+91.4\% improvement in tail-trend detection precision@500} and \hyperref[sec:AMD]{\textbf{+19\% query generation accuracy}} over industry baselines, while sustaining stable performance after multi-week online training. This work demonstrates that \textbf{LLM-generated synthetic search signals}, when aligned and continually updated, unlock timely trend understanding in low-traffic search environments.
\end{abstract}

\ccsdesc[500]{Information systems~Content ranking}

\keywords{Trending news prediction, Synthetic query generation, LLM, Continual learning, Preference optimization, Mix-DPO}
\maketitle
\section{Introduction}
Trending query detection in social media search is critical and can be used in many cases. It enhances ranking quality by giving more accurate freshness sensitivity signal. Traditional search trend detection methods rely on query volume or keyword frequency, e.g., Google Trends uses statistical heuristics over massive query logs, and X applies Poisson models to detect spikes in keyword and hashtag usage \cite{mathioudakis2010twittertrends}. Platforms with low search traffic struggle to surface emerging or tail trends using query signals alone.

We tackled the trend detection problem from a new perspective: rather than waiting for trending query volume to emerge, we use LLM to generate trending queries from newly created posts. We propose \textbf{RTTP} (Real-Time Trend Prediction), which leverages a continually updated LLM (CL-LLM) to generate search-style queries from news posts and score using both creators' quality and engagement strength. Deeper interaction signals (e.g., comments over likes) help surface trends with greater granularity and timeliness.

CL-LLM is designed for regular adaptation, relying on continual training to be aligned with evolving topics. To ensure alignment stability, we introduce a \textbf{Mix-Policy DPO} training strategy that blends on-policy and off-policy data, addressing the ``squeezing effect'' \cite{yi2024squeezing} that training aggressively pushes down the probability of less-preferred responses, causing probability mass to be funneled into the single most likely correct answer.
\section{Related Work}
\subsection{Industrial Approaches to Trend Detection}
Platforms such as Google, X (formerly Twitter)  and Amazon have developed large-scale trend detection systems. Google Trends relies on statistical heuristics on billions of search queries \cite{djorno2025restoringforecastingpowergoogle}, X uses Poisson models for token bursts \cite{mathioudakis2010twittertrends}, and Amazon employs time-series models including InceptionTime with engagement signals \cite{gayatri2023trendspotter}. Traditional methods depend on query volumes and often fail to capture emerging or long-tail trends in low-traffic environments.

\subsection{LLMs for Trend Understanding and Generation}
LLMs are increasingly applied to trend detection in sparse-data scenarios. Domain adaptation techniques \cite{parmar2024reuse} align models with evolving social content, but static LLMs struggle to track real-time knowledge changes. Our CL-LLM enables adaptive trend understanding through continual updates.

\subsection{LLM Continual Learning}
Static LLM without latest world knowledge can't catch up social media users' interest shifting. User Preference alignment is the goal to optimize LLM responses \cite{bai2025copo}. 
LLM continual learning techniques—including replay, LoRA, and learning rate scheduling—seek to maintain the stability of LLM \cite{shi2023unified, hu2021lora}. However, catastrophic forgetting remains a key challenge \cite{li2024examining,luo2025empirical}. Domain-adaptive pretraining (DAP) can mitigate some issues but typically requires large curated datasets.
Direct Preference Optimization (DPO) \cite{rafailov2023dpo} learns from pairwise preferences, but off-policy DPO suffers from the \emph{squeezing effect}, reducing reasoning diversity. Step-DPO \cite{xu-etal-2025-full} and Mix-Policy DPO balance alignment and stability, especially with frequent SFT.

\section{RTTP System}

\subsection{System Architecture}
The RTTP system takes newly created posts and engagement signals as input and outputs a ranked list of predicted trending queries. The pipeline consists of:

\begin{itemize}
    \item \textbf{Triggering upon post creation:} Creators publish posts that gather interactions from creators' followers. Deep engagement (comments, reshares) and shallow engagement (likes, reactions) are distinguished.
    \item \textbf{Query Generation and Parsing:} The CL-LLM parse latest posts and generates search queries, and assigns location metadata etc. We use LLaMA 3.3 70B continual training on real user queries.
    \item \textbf{Scoring \& Ranking by User Engagement:} Queries are aggregated and weighted by engagement and content authority. Deeper engagement and authoritative sources receive higher weights.
\end{itemize}

Trend scores are computed as:
\begin{small}
\begin{align}
\text{CreatorAuthority} &= \log(\text{FollowerCount}) + \sum_i \text{AuthoritySignal}_i,\\
\text{TrendingScore} &= \text{CreatorQuality} + \sum_i w_i \cdot E_i, 
\end{align}
\end{small}
\textit{FollowerCount} represents the total number of followers for a facebook account, \textit{$\text{AuthoritySignal}_i$} represents facebook account authority features, such as if a facebook account is  Meta Verified, add a reward score to \textit{CreatorAuthority}. \\
Given a post, an user has different types of engagement, for example reacting with a like/surprise emoji, re-sharing the post, commenting etc. $w_i$ denotes the weight for type $i$ user engagement, $E_i$ denotes the amount of type $i$ engagement. For example, we give commenting higher weight than clicking an emoji. 

\subsection{CL-LLM}
CL-LLM plays a central role in generating search queries, where user behavior shifts rapidly over time. Static LLMs struggle to adapt to emerging knowledge and evolving user preferences, leading to degraded relevance in production search systems. In contrast, CL-LLM is continually updated to incorporate new information while mitigating \emph{catastrophic forgetting} \cite{luo2025empirical} issue, enabling it to maintain both adaptability and long-term retention.

\begin{figure}[!ht]
    \centering
    \includegraphics[width=0.45\textwidth]{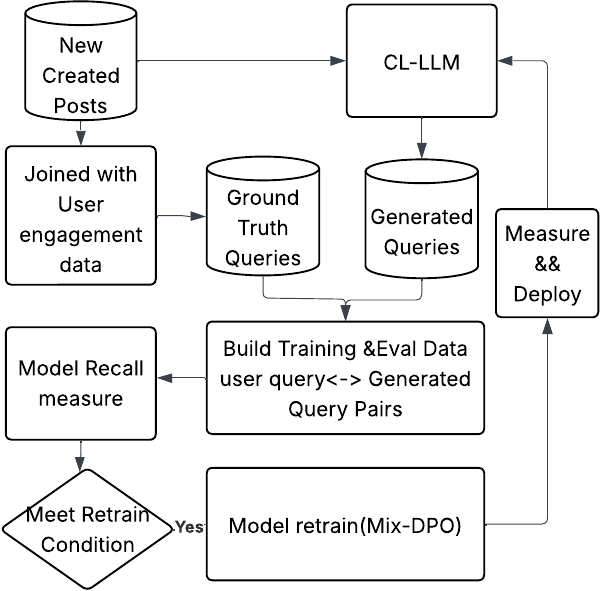}
    \caption{CL-LLM continual training workflow.}
    \label{fig:cl-llm-workflow}
\end{figure}

\subsubsection{Mix-Policy DPO}
Mix-Policy DPO mitigates catastrophic forgetting by jointly leveraging both on-policy and off-policy data throughout optimization. 
From search query-click log, we treat the real user queries associated with posts as the ground-truth queries. On-policy training data is constructed from cases where CL-LLM successfully generates queries that align with this ground truth. In contrast, off-policy data consists of newly emerging posts for which the model fails to produce the ground-truth queries, often due to topic novelty or distribution shift. For off-policy examples, the ground-truth queries are used as positives, while the model’s mismatched predictions are treated as negatives.

On-policy samples preserve the model’s existing capabilities by continually reinforcing behaviors it already performs well. Meanwhile, off-policy samples introduce emerging or previously mispredicted content, which drives adaptation to new distributions without overwriting prior knowledge. By maintaining a balanced learning signal across stable and novel domains, the mix-policy approach reduces the tendency of the model to overfit to only the latest data, thereby sustaining long-term reasoning stability and performance retention under real-time trend shifts. 
Let $x$ be a prompt, and $y_\text{win}, y_\text{lose}$ preferred and non-preferred completions:

\begin{small}
\begin{equation}
L = -\log \sigma \Big( \beta [ \log \frac{\pi_\theta(y_\text{win}|x)}{\pi_\text{ref}(y_\text{win}|x)} - \log \frac{\pi_\theta(y_\text{lose}|x)}{\pi_\text{ref}(y_\text{lose}|x)} ] \Big)
\end{equation}
\end{small}

Throughout the experimental and evaluation phases, the off-policy to on-policy ratio $\rho = 1{:}9$ avoids the squeezing effect. Retraining is triggered if the query generation recall metric regresses below a predefined threshold.

\subsubsection{Training flow}
As Figure 1, CL-LLM takes post titles, body content to generate:
\begin{itemize}
    \item Search queries reflecting post topics and user query patterns.
    \item Location metadata (country/state).
\end{itemize}

CL-LLM retraining is triggered based on \texttt{Recall@3} which is defined as: a post $P$ has ground-truth query $Q$ if a user searches $Q$ and engages with $P$. If $Q$ appears in the top 3 generated queries, recall is successful. Retraining is triggered when \texttt{Recall@3} drops by more than a threshold, like 10\% in evaluation.

\section{Experiments}
Three evaluation tasks are conducted: a) trend prediction evaluation. b) query generation accuracy evaluation, the key component in RTTP. c) CL-LLM model stability evaluation after rounds of continual training. 

\subsection{Datasets}

\textbf{Search Trending Queries.}
To evaluate the real-world trend detection capabilities, we utilized a dataset comprising one month of search logs and user engagement from Facebook, alongside human-annotated labels curated by internal domain experts. We acknowledge that the proprietary nature of this data introduces inherent constraints on external reproducibility. However, this reliance on internal logs is necessary to provide a high-fidelity benchmark of actual user behavior at scale. To ensure ethical data usage, all processing adhered to strict privacy and governance protocols; data was de-identified, aggregated, and handled in accordance with internal compliance standards \\
\textbf{Reasoning Data.}  To assess the effectiveness of the Mix-Policy DPO strategy and our accuracy-decay-based continual training approach, we employed the MMLU benchmark \cite{hendrycks2021mmlu}. Because RTTP is specifically designed for text-based trend understanding, we focused our evaluation on linguistic and general knowledge categories.

\subsection{Trend Prediction Evaluation}

\textbf{Evaluation Design} 
We evaluated industry solutions using precision as metrics. For each day, using all methods identified trending queries and ranked candidates based on search volume. All detected trending queries were reviewed by internal rating expert to calculate the precision.

The baselines include:  
\begin{itemize}
    \item Poisson model on search traffic data: a Poisson-based model proposed by X \citep{mathioudakis2010twittertrends},  similar to Google Trends that detects spikes in query search volume.
    \item Poisson model on search traffic data + generated queries: augments baseline with synthetic queries from the CL-LLM.
    \item RTTP: combines generated queries with engagement-weighted scoring.  Engagement is defined as users reacting to post bys selecting pre-defined emotions, like "Love", "Haha" etc
\end{itemize}

\begin{figure}[!ht]
    \centering
    \includegraphics[width=0.45\textwidth]{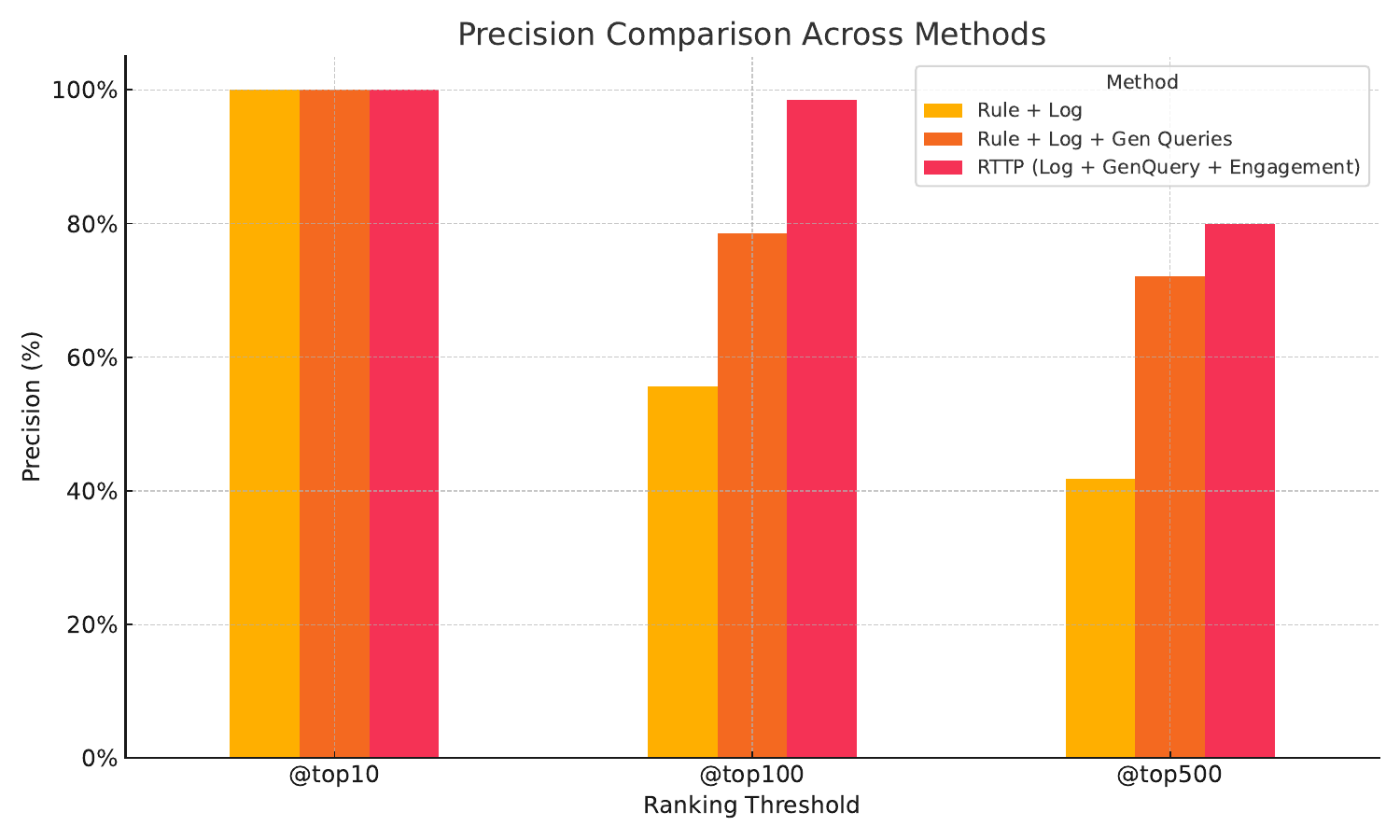}
    \caption{Precision across ranking tiers for trend detection methods.}
    \label{fig:precision}
\end{figure}

Figure~\ref{fig:precision} shows precisions across three methods. Precision generally drops for lower-volume trending queries. RTTP consistently improves precision at all ranking tiers. Notably, in the top 500 trending queries per day, precision@500 increases from 41.8\% to 80\%, a +91.4\% relative improvement($P$-value< 0.001) given 300 data points from 1 month, highlighting the benefit of incorporating both generated queries and engagement weighting.

\subsection{Query Generation Accuracy Evaluation} 
The core component of RTTP is generating queries from content. Therefore, we evaluated accuracy of CL-LLM model. 
\subsubsection{Accuracy Metrics Definition}\label{sec:AMD}
Given a post $P$, if users searched a query $Q$ and engaged with $P$, $Q$ is considered a ground-truth query. Posts always have multiple ground-truth queries. If any ground-truth query appears among the top K queries generated by the CL-LLM, it counts as a successful generation. We use K = 3 after empirical tuning. 

\subsubsection{Evaluation Design}

We generated queries hourly from the most recently created content and evaluated model performance using the metrics described above. 

As baselines, we evaluated standard industry continual-training strategies, including replay-based SFT with varying replay ratios \cite{ibrahim2024continual} and learning-rate scheduling approaches \cite{parmar2024reuse}. As shown in Figure 3, Mix-DPO initially exceeded all SFT-based baselines by roughly 4\%. Over time, the gap widened: after one week, the accuracy of all SFT-based models dropped to approximately 76\%, while Mix-DPO maintained around 90.5\% accuracy(19\% improvement). This highlights Mix-DPO’s superior robustness and its ability to preserve performance during continual updates.
The results suggest two key exploration directions: (a) identifying scenarios where Mix-DPO outperforms SFT, and (b) analyzing the causes of performance differences. Deep dives revealed: SFT-based solutions primarily learn straightforward patterns, e.g., using words in titles or first sentences. Mix-DPO-generating queries can better leverage world knowledge beyond post content. 
For instance, when processing a post titled \textit{"Mounts of Mayhem"}—which does not explicitly mention \textit{"Minecraft"}—Mix-DPO is capable of generating \textit{"Minecraft"} as a relevant query by leveraging latent associations. In contrast, SFT-based solutions are often limited to extractive keyword identification from the provided text. It is well-documented that extensive Supervised Fine-Tuning (SFT) can erode the embedded knowledge within Large Language Models (LLMs). Our results demonstrate that Mix-DPO more effectively preserves the model's reasoning capabilities and pre-trained knowledge, whereas increased SFT iterations tend to accelerate the degradation of these existing competencies
\begin{figure}[!ht]
    \centering
    \includegraphics[width=0.45\textwidth]{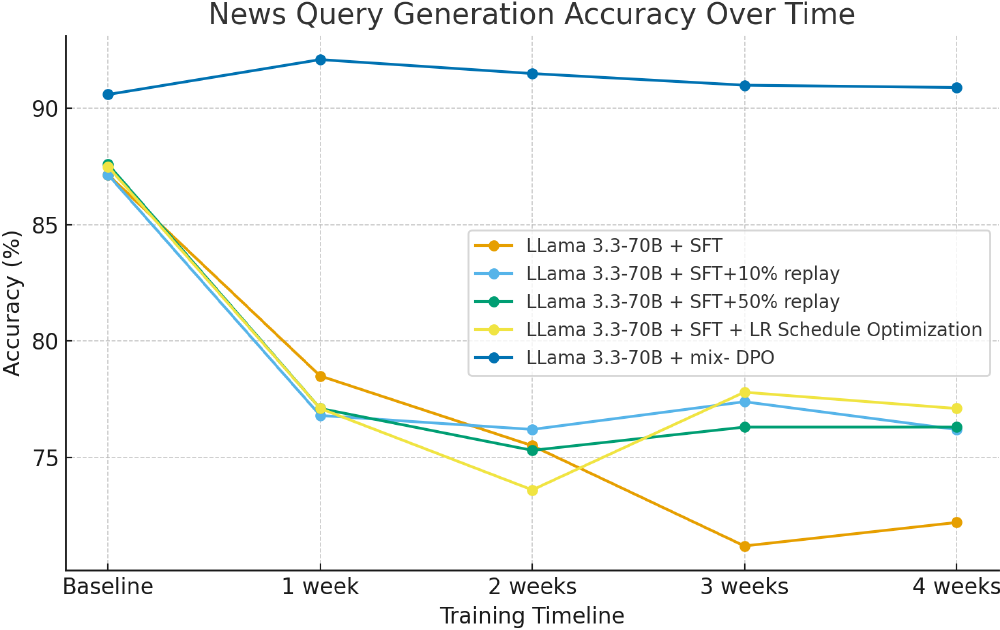}
    \caption{Query generation accuracy over 4 weeks. Mix-DPO maintains higher accuracy compared to SFT-based methods.}
    \label{fig:query-accuracy}
\end{figure}
\subsection{Reasoning Stability Evaluation}
\begin{figure}[!ht]
    \centering
    \includegraphics[width=0.45\textwidth]{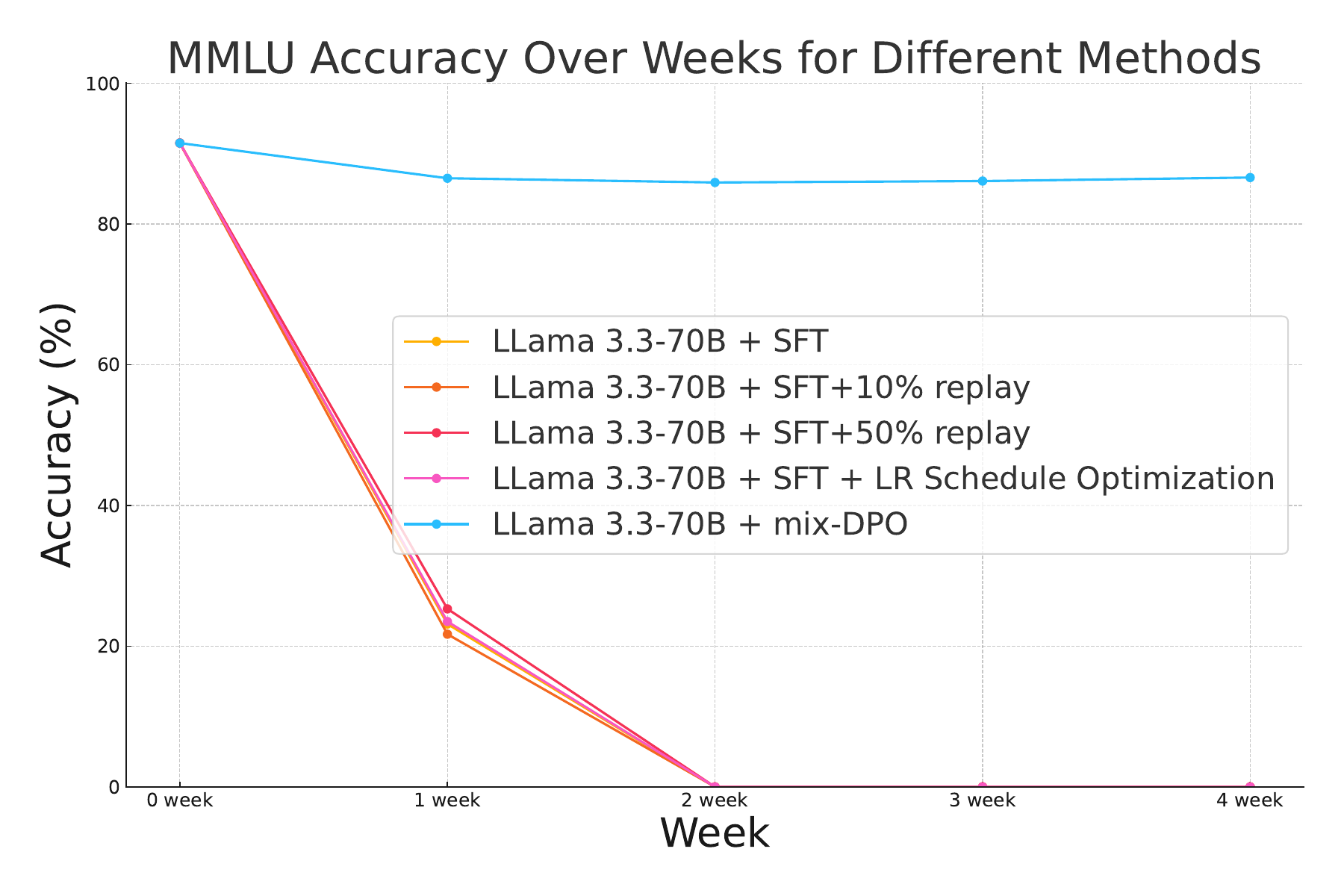}
    \caption{MMLU benchmark accuracy over a month for continual training. Mix-DPO maintains reasoning capability }
    \label{fig:mmlu-stability}
\end{figure}

The baseline setup follows Section 4.3, with the evaluation set changed from query generation to the MMLU benchmark. Since RTTP primarily targets text trend understanding, we
excluded categories of coding and mathematics.
Figure~\ref{fig:mmlu-stability} shows the metrics for all SFT-based models quickly drop to near 0 after 2 rounds of 500 steps(1000 batch-size) continual training. In contrast, Mix-DPO exhibits about 5\% drop on the MMLU set and maintains this performance throughout a month.
Further analysis reveals several causes for the degradation of reasoning capabilities in SFT-based models:
\begin{itemize}
    \item \textbf{Divergent experimental setups.} Prior works such as \citet{ibrahim2024continual} and \citet{parmar2024reuse} assume continual pretraining under data distributions similar to the pretraining corpus. LLaMA 3.3-70B, trained on 15T tokens, requires large volumes of high-quality query instances for retraining. In production, GPU constraints limit query instances per retraining cycle, forcing aggressive sampling. This mismatch in scale and data quality contributes to faster degradation of reasoning ability.
    
    \item \textbf{Unstable data quality from user-generated content.} Unlike structured sources like Wikipedia etc, social media posts are informal and diverse. The token distribution diverges significantly from the pretraining corpus, contributing to performance degradation, as observed by \citet{li2024examining}.
    
    \item \textbf{Post-aligned model constraints.} Unlike prior continual pretraining work operating on pretrained checkpoints \cite{parmar2024reuse}, LLaMA 3.3-70B is a post-trained (instruction-aligned) model. Experiments indicate that SFT-aligned models lose instruction-following capability rapidly, with near-total degradation after just 400 steps.
\end{itemize}

\section{Conclusion}
This paper proposes RTTP system which generates and weights synthetic queries using a continually updated LLM (CL-LLM). At its core is the Mix-Policy DPO training strategy, which combines on-policy and off-policy preference data to mitigate catastrophic forgetting while preserving reasoning capabilities. RTTP is verified with real user data of 
Facebook search and Meta AI app platforms. 

\bibliographystyle{ACM-Reference-Format}
\balance
\bibliography{rttp_full_references.bib}

\end{document}